# GAMMA RADIATION MEASUREMENTS AND DOSE RATES IN COMMONLY USED BUILDING MATERIALS IN CYPRUS

**F. Michael[1], Y. Parpottas[2], H. Tsertos[1]**

[1]Department of Physics, University of Cyprus, P.O. Box 20537, 1678 Nicosia, Cyprus

[2]School of Engineering and Applied Sciences, Frederick University, 1036 Nicosia, Cyprus

## ABSTRACT

A first comprehensive study is presented on radioactivity concentrations and dose rates in 87 commonly used materials, manufactured or imported in Cyprus, for building purposes. The natural radioactivity of $^{40}K$, $^{232}Th$, $^{238}U$ and $^{226}Ra$ is determined using high-resolution gamma ray spectroscopy. The respective dose rates and the associated radiological effect indices are also calculated. A comparison of the measured specific activity values with the corresponding world average values shows that most of them are below the world average activity values. The annual indoor effective dose rates received by an individual from three measured imported granites and four measured imported ceramics are found to be higher than the world upper limit value of 1 mSv y$^{-1}$. Hence, these materials should have a restricted use according to their corresponding calculated activity concentration index values and the related EC 1999 guidelines.

## INTRODUCTION

The most important naturally occurring radionuclides in building materials from rocks and soils are the radioactive isotope of potassium ($^{40}K$) and the radionuclides from thorium ($^{232}Th$) and uranium ($^{238}U$) decay series. It is of great interest to determine the population exposure to radiation from such materials since humans spend 80% of their time indoors [1]. In Cyprus, a series of previous studies were conducted by our laboratory with the objective to determine the radioactivity levels and associated dose rates from surface soils and some imported granites [2-5]. In this study, we extent our measurements to cover, systematically, nearly all the primary (raw) materials used largely in Cyprus from the building and construction industry. For this purpose, the specific activities of those naturally occurring radionuclides were measured in 87 representative samples, by means of high-resolution gamma-ray spectroscopy. The respective dose rates and the associated radiological effect indices are also reported and compared with the world median values and the accepted upper limits.

## METHODOLOGY

### Samples

A total of 87 samples, used for building purposes in Cyprus, were collected: 29 samples from domestic quarries, 13 raw cement samples from domestic industries, 11 clay samples from brick and tile industries, 8 imported granite rock samples, 10 imported marble rock samples, 13 imported ceramic floor-tile samples, and 3 domestic and imported decorative stones. First, the samples were crushed and milled to a fine powder. Then, they were dried in an oven at 100 $^{\circ}$C for 24 hours to ensure that moisture was completely removed. Last, they were sieved through a 1 mm mesh, and hermetically sealed in standard 1000 ml plastic Marinelli beakers with bore diameter of 85 mm. They were stored for about four weeks to ensure secular equilibrium between $^{226}$Ra and $^{222}$Rn and their decay products.

### Measurements

A coaxial cylinder (55 mm in diameter and 73 mm in length) p-type high purity germanium (HPGe) detector, with an efficiency of 30% relative to a 3″x 3″ NaI(Tl) scintillator, and an energy resolution (FWHM) of 1.8 keV for the 1.33 MeV reference transition of $^{60}$Co, was utilized for the measurements. The prepared Marinelli beakers were placed on the detector endcap. Both the sample and the detector were surrounded by a cylindrical graded-Z shield of 5-cm thickness of lead, 1-cm thickness of iron and 1-cm thickness of aluminum to suppress the background radiation. A spectroscopic amplifier, with an efficient pile-up rejector, and an 8k ADC (Analog-to-Digital Converter) processed the signal. The MAESTRO-32 multi-channel analyzer emulation software was utilized for data acquisition, storage, display and online analysis of the spectra. Each sample was measured for 24 hrs to obtain good statistics. Measurements with an empty Marinelli beaker, under identical conditions, were also carried out to determine the ambient background in the laboratory site. The latter was subtracted from the measured spectra to obtain the net radionuclide activities. The activity concentrations of $^{40}$K, $^{232}$Th and $^{238}$U from the reference soil sample IAEA-326 was initially measured, under identical conditions, for quality assurance purposes.

The gamma-rays of interest ranged from 50 - 3000 keV. The detector energy-dependent efficiency was determined using a calibrated [152]Eu gamma reference source [6]. The latter is sealed in a standard Marinelli beaker with an active volume of 1000 ml and a bore diameter of 85 mm. The source average density is 1 g cm[-3] and the source initial reported activity was 10.1 kBq/kg with an uncertainty of 3 %. The uncertainty in the calculated efficiency was estimated to be 5 %.

Offline analysis was performed using the Gamma Vision-32 software. All statistically significant photopeaks in the spectrum were simultaneously fitted. Menu-driven reports were generated with information about the channel centroid, the energy, the counts of the net area and the background, the intensity and width of the identified and unidentified photopeaks as well as the photopeak activity and the average activity for each detected radionuclide.

## Calculations

### Activity concentrations

The activity concentrations for the [40]K, [232]Th, [238]U and [226]Ra radionuclides were calculated using the detected photopeaks in the spectra. Since secular equilibrium was reached between [232]Th and [238]U and their decay products, the [232]Th activity concentration was determined from the weighted average activity concentrations of the [228]Ac detected photopeaks at 911.07 and 968.90 keV, and the [238]U activity concentration was determined from the weighted average activity concentrations of the [214]Bi detected photopeaks at 609.32 and 1764.51 keV. The 1460.75 keV gamma-ray of [40]K and the 185.99 keV gamma ray of [226]Ra were used to determine the activity concentrations of the corresponding radionuclides. In the present study, the activity concentration of the [226]Ra radionuclide in the [238]U decay series was not used for the activity concentration of [238]U radionuclide because it may have slightly different concentration than the [238]U radionuclide. This is due to the separation that may occur between its parent [230]Th and [238]U and also due to the fact that [226]Ra has greater mobility in the environment [1].

Hence, the activity concentration ($A_{E,i}$) in Bq kg[-1], for a radionuclide i with a detected photopeak at energy E, is obtained from the following equation:

$$A_{E,i} = \frac{N_{E,i}}{\varepsilon_E \times t \times \gamma_d \times M} \qquad (1)$$

where $N_{E,i}$ is the net peak-area of the radionuclide i at energy E, $\varepsilon_E$ is the detector energy-dependent efficiency at energy E, t is the counting live time in sec, $\gamma_d$ is the gamma-ray yield per disintegration of the nuclide i for its transition at energy E, and M is the mass of the sample measured in kg.

The total uncertainty in the activity concentration was calculated, as follows [7]:

$$\sigma_{tot} = \sqrt{\sigma_{st}^2 + \frac{1}{3}\sigma_{sys}^2} \qquad (2)$$

where $\sigma_{st}$ is the counting statistical error and $\sigma_{sys}$ is the total systematic error that comes from the uncertainty in the efficiency fitting function (1-10%), the uncertainty in the activity of the $^{152}$Eu gamma reference source (3%), and the uncertainty in the nuclide master library (1-2%).

*Absorbed dose rates*

The absorbed dose rate (D) in nGy/h from the $^{40}$K isotope and the radionuclides from $^{232}$Th and $^{238}$U decay series, was calculated as follows [8,9]:

$$D = 0.52813\, A_{Th} + 0.38919\, A_U + 0.03861\, A_K \qquad (3)$$

where $A_{Th}$ is the mean activity concentration of $^{232}$Th in Bq/kg, $A_U$ is the mean activity concentration of $^{238}$U in Bq/kg, and $A_K$ is the mean activity concentration of $^{40}$K in Bq/kg. The calculation refers to the dose rate, in air at a height of 1.0 m above the ground, if the naturally occurring radionuclides are uniformly distributed. The activity concentrations of the three radionuclides were multiplied by the corresponding Dose Rate Conversion Factors (DRCF) which are the absorbed dose rates in air per unit activity per unit of soil mass, in nGy/h per Bq/kg. In the present work, the DRCF's are the average values obtained from the corresponding values derived by three Monte Carlo codes [10-12]: the Los Alamos MCNP code (1986), the GEANT code from CERN (1993), and the MC Monte Carlo code developed by the Nuclear Technology Laboratory of the Aristotle University of Thessaloniki, Greece.

*Annual effective dose rates*

The annual outdoor effective dose rate ($D_{E,\,out}$) to a member of the population is calculated from the absorbed dose rate taking into account the conversion factor (CF) of the absorbed dose in air to the corresponding effective dose, and the occupancy factor ($OF_{out}$). It is equal to:

$$D_{E,out} = D \times CF \times OF_{out} \qquad (4)$$

where $D_{E,out}$ units are in Sv/y, D units are in Gy/h, CF = 0.7 Sv/Gy [1,13], and OF = f × 24 hrs × 365.25 days. Since humans are expected to spend 20% of their time outdoors and 80% indoors [1,13], $f_{out} = 0.2$ and $f_{in} = 0.8$, for the annual outdoor and indoor effective dose rates, respectively.

It has been found by previous extensive investigations of soil samples, which were collected from the main island bedrock surfaces [4], on the one hand, and *in situ* high-resolution gamma spectrometry in urban areas [14], on the other hand, that the results obtained outdoors on the radionuclide activities and on the corresponding associated gamma dose rates are very similar [14]. This implies that the construction and building materials used in urban areas do not affect the outdoor gamma dose rate, revealing that they are mostly of local origin. In the interior of buildings and dwellings (indoor environments), where the people spend most of their time (about 80%), however, the situation becomes more complex: the gamma-source geometry is generally unknown, and gamma radiation comes not only from the ground, but also from the walls and roofs. The building materials act as sources of radiation and also as shields against outdoor radiation. Therefore, *in situ* high-resolution gamma spectrometry was used to correctly determine the indoor gamma dose rate. It has been found from such measurements that the ratio (R) of indoor/outdoor gamma dose rate to be 1.4 ± 0.5 [14], which has been used in the present work to determine realistically the indoor gamma dose rate in a typical Cyprus house.

Hence, the annual indoor effective gamma dose rate is equal to:

$$D_{E,in} = D \times CF \times OF_{in} \times R \qquad (5)$$

A significant source to the indoor dose rates (56.4%) is coming from the radon concentration [15]. Such systematic indoor airborne radon concentration measurements [15] were conducted to determine realistically the indoor radon dose rates in a typical Cyprus house.

*Radium equivalent activity*

The radium equivalent activity ($Ra_{eq}$) in Bq/kg was introduced to define uniformity in respect to radiation exposure [13,16,17]. This is because the $^{226}Ra$, $^{232}Th$ and $^{40}K$ distribution in building materials is not uniform [18]. The radium equivalent index compares the activity concentration of materials containing different amounts of these radionuclides. It is the weighted sum of $^{226}Ra$, $^{232}Th$ and $^{40}K$ activity concentrations, based on the assumption that 10 Bq/kg of $^{226}Ra$, 7 Bq/kg of $^{232}Th$ and 130 Bq/kg of $^{40}K$ produce the same gamma dose rate. It is calculated as follows [13]:

$$Ra_{eq} = A_{Ra} + 1.43A_{Th} + 0.077A_K \qquad (6)$$

where $A_{Ra}$, $A_{Th}$ and $A_K$ are the mean activity concentrations of $^{226}Ra$, $^{232}Th$ and $^{40}K$ in Bq/kg, respectively.

*External hazard index*

The model of the external hazard index ($H_{ex}$) places an upper limit to the external gamma radiation dose from building materials to unity, which corresponds to a radium equivalent activity of 370 Bq/kg. It is defined [16,18] as:

$$H_{ex} = \frac{A_{Ra}}{370 \text{ Bq/kg}} + \frac{A_{Th}}{259 \text{ Bq/kg}} + \frac{A_K}{4810 \text{ Bq/kg}} \leq 1 \qquad (7)$$

where $A_{Ra}$, $A_{Th}$ and $A_K$ are the mean activity concentrations of $^{226}Ra$, $^{232}Th$ and $^{40}K$ in Bq/kg, respectively.

The value of this index should be less than unity to keep the radiation hazard negligible [19].

*Activity concentration index*

The activity concentration index (I) was derived for identifying whether the European Commission [6] guidelines about building material usage are met.

It is defined as follows:

$$I = \frac{A_{Ra}}{300 \text{ Bq/kg}} + \frac{A_{Th}}{200 \text{ Bq/kg}} + \frac{A_K}{3000 \text{ Bq/kg}} \qquad (8)$$

where $A_{Ra}$, $A_{Th}$ and $A_K$ are the mean activity concentrations of $^{226}Ra$, $^{232}Th$ and $^{40}K$ in Bq/kg, respectively. According to these guidelines [6], the building materials that increase the annual outdoor effective dose received by an individual an amount of 0.3 mSv should not be considered as hazardous. On the other hand, building materials that increase the annual outdoor effective dose received by an individual an amount more than 1 mSv/y should be taken into account from radiation protection point of view. The activity concentration index values that correspond to the 0.3 mSv exemption criterion and the 1 mSv upper limit, for materials used in bulk amounts (e.g. concrete), are 0.5 and 1, respectively. The corresponding activity concentration index values for superficial and other materials with restricted use (e.g. tiles) are 2 and 6, respectively.

**RESULTS AND DISCUSSION**

The measured activity concentrations of $^{226}$Ra, $^{238}$U, $^{232}$Th, and $^{40}$K for the 87 collected samples are shown in Table 1. The activity concentration of samples coming from the same region were averaged and presented in the Table. In some cases, we reported the minimum detectable activity (MDA) since the activity concentrations were below that limit.

The $A_{Ra}$, $A_U$, $A_{Th}$, $A_K$ in the samples collected from domestic quarries range from 0.1 – 40.2 Bq/kg, 0.8 – 21.2 Bq/kg, 0.3 – 6.4 Bq/kg, and 4.6 – 147.3 Bq/kg, respectively. The $A_{Ra}$, $A_U$, $A_{Th}$, $A_K$ in the domestic cement samples range from 4.4 – 60.3 Bq/kg, 3.8 – 24.5 Bq/kg, 0.6 – 12.3 Bq kg$^{-1}$, 4.1 – 289.2 Bq/kg, respectively. The $A_{Ra}$, $A_U$, $A_{Th}$, $A_K$ in the domestic clay samples range from 0.1 - 33.8 Bq/kg, 2.7 – 20.6 Bq/ kg, 1.8 – 16.0 Bq/kg, 59.3 – 376.9 Bq/kg, respectively. The $A_{Ra}$, $A_U$, $A_{Th}$, $A_K$ in the imported granite samples range from 0.2 – 81.0 Bq/kg, 3.2 – 107.0 Bq/kg, 5.7 – 260.0 Bq/kg, 920.7 - 1576. Bq/kg, respectively. The $A_{Ra}$, $A_U$, $A_{Th}$, $A_K$ in the imported marble samples range from 0.1 – 79. Bq/kg, 1.3 –35.3 Bq/kg, 0.1 – 2.1 Bq/kg, 0.2 – 65.6 Bq/kg, respectively. The $A_{Ra}$, $A_U$, $A_{Th}$, $A_K$ in the imported ceramic samples range from 30.1 – 299.5 Bq/kg, 8.6 - 140. Bq/kg, 1.7 - 208.0 Bq/kg, 4.1 - 1051.8 Bq/kg, respectively. Last, the $A_{Ra}$, $A_U$, $A_{Th}$, $A_K$ in the decorative stone samples range from 23.4 – 53.3 Bq/kg, 14.3 – 20.1 Bq/kg, 1.1 – 50.8 Bq/kg, 5.8 – 1010.4 Bq/kg, respectively.

The reported world median radioactivity levels for $^{226}$Ra, $^{238}$U, $^{232}$Th, $^{40}$K are 35, 35, 30, and 400 Bq kg$^{-1}$, respectively[1]. In this work, 24 samples are found to have higher $^{226}$Ra activity concentration values than the world median radioactivity levels (from quarries: 1, cements: 4, granites: 3, marbles: 4, ceramics: 10, decorative stones: 2) , 16 samples have higher $^{238}$U activity concentration values than the world median radioactivity levels (granites: 4, ceramics: 12), 18 samples have higher $^{232}$Th activity concentration values than the world median radioactivity levels (granites: 5, ceramics: 12, decorative stones: 1) , and 19 samples have higher $^{40}$K activity concentration values than the world median radioactivity levels (granites: 7, ceramics: 11, decorative stones: 1).

In general, the activity concentrations in the imported granites (7 out of 8) and ceramics (12 out of 13) are higher than the corresponding world median values. On the other hand, the measured samples from the domestic quarries, the domestic cements and clays, and the imported marbles present low radioactivity levels. Figure 1 presents the $A_{Ra}$, $A_U$, $A_{Th}$, $A_K$ of the measured samples. The horizontal solid line in each graph corresponds to the world median radioactivity level. The vertical dashed lines in each graph correspond to the seven sample categories in Table 1.

Table 2 lists the calculated total absorbed dose rates, the annual outdoor and indoor effective dose rates, the radium equivalent activity, the external hazard index, and the activity concentration index values for the corresponding measured samples in Table 1. The total absorbed dose rates of the measured samples range from 0.6 − 225.8 nGy/h. The corresponding world median value is 60 nGy/h [1]. Figure 2 shows the frequency distribution of the total absorbed dose rates for the measured samples of Table 1. Table 2 and Figure 2 show that 71% of the measured samples have absorbed dose rate values lower than the corresponding world median one, while the respective values from 20 measured samples are higher than the reported median one. The total absorbed dose rates are also plotted in the top-left panel of Figure 3. The horizontal solid line corresponds to the world average value (60 nGy/h) and the vertical dashed lines correspond to the seven sample categories as those were described in Table 1. Most of the calculated absorbed dose rate values from the measured granites and ceramics are higher than the world average one. The same occurs with the annual indoor effective dose rates of those granites and ceramics, as can be seen at the top-right panel of Figure 3. The horizontal solid line on the plot corresponds to the world median value of the indoor effective dose rate (410 μSv/y) [1].

A comparison of the calculated radium equivalent index from the measured samples with the corresponding world accepted upper limit ($Ra_{eq}$ = 370 Bq/kg) [13,18], and a comparison of the calculated external hazard index from the measured samples with the corresponding world accepted upper limit ($H_{ex} = 1$) are shown in the graphs (bottom-left and bottom-right panels) of Figure 3, respectively. The horizontal solid lines correspond to the abovementioned upper limits and the vertical dashed lines correspond to the seven sample categories in Table 1. The calculated radium equivalent index and the external hazard index values from three measured imported granites and four measured imported ceramics are above the corresponding UNSCEAR upper limits. Averaging, separately, the $A_{Ra}$, $A_{Th}$, $A_K$ of those three granites, and taking into account Eq. 6, we calculated the relative contribution of $^{226}Ra$ , $^{232}Th$, and $^{40}K$ to $Ra_{eq}$ to be 25.7 %, 46.8%, 27.4%, respectively. The same calculation was repeated for those four ceramics. In this case, the relative contribution of $^{226}Ra$, $^{232}Th$, and $^{40}K$ to $Ra_{eq}$ is 43.7%, 45.8%, and 10.4%, respectively.

The activity concentration index values are also listed in Table 2. They range from 0.003 – 1.974. Only four granites and six ceramics are found to have activity concentration index values between 1 and 2. According to the EC guidelines [6], materials should be used in bulk amounts if their activity concentration index values do not exceed the upper limit of 1. Further, materials with restricted use should be exempted from any criteria if their activity concentration index values are lower than 2.

**CONCLUSIONS**

High-resolution gamma ray spectroscopy was used to determine the natural radioactivity concentration of $^{40}K$, $^{232}Th$, $^{238}U$ and $^{226}Ra$ presented in 87 domestic (raw and rock) or imported (rock) materials, commonly used for building purposes in Cyprus. The total absorbed dose rates and the annual indoor effective dose rates were calculated from the activity concentrations and compared to the corresponding world median values. The annual indoor effective dose rates received by an individual from three imported granites and four imported ceramics are found to be higher than the world upper limit value of 1 mSv/y. Hence, a restricted use of those materials is recommended, according to the EC 1999 guidelines.

ACKNOWLEDGEMENTS

We would like to thank all directors of the quarries, companies, and industries who provided to us the samples. Special thanks go to P. Demetriades and M. Tzortzis from the Radiation Protection and Control Services, of the Department of Labor Inspection of the Ministry of Labor and Social Insurances, for the permanent and fruitful collaboration. We also appreciate the help from the Department of Geological Survey, of the Ministry of Agriculture, Natural Resources and Environment, and the help from the Department of Public Works, of the Ministry of Communications and Works, in crushing the samples.

## REFERENCES

1. UNSCEAR. *Sources and Effects of Ionizing Radiation*. Report to General Assembly, Vol. 1, Annex B, United Nations, New York (2000).

2. Tzortzis, M., Tsertos, H., Christofides, S. and Christodoulides, G. *Gamma-ray measurements of naturally occurring radioactive samples from Cyprus characteristic geological rocks*. Radiat. Meas. **37**, 221-229 (2003).

3. Tzortzis, M., Tsertos, H., Christofides, S. and Christodoulides, G. *Gamma radiation measurements and dose rates in commercially-used natural tiling rocks (granites)*. J. Environ. Radioact. **70**, 223-235 (2003).

4. Tzortzis, M., Svoukis, E. and Tsertos, H. *A comprehensive study of natural gamma radioactivity levels and associated dose rates from surface soils in Cyprus*. Radiat. Prot. Dosim. **109**(3), 217-224 (2004).

5. Tzortzis, M. and Tsertos, H. *Determination of thorium, uranium and potassium elemental concentrations in surface soils in Cyprus*, J. Environ. Radioact. **77**(3), 325-338 (2004).

6. European Commission Report. *Radiological Protection Principles concerning the Natural Radioactivity of Building Materials*. Radiation Protection 112 (1999).

7. EG&G ORTEC. *Gamma Vision 32: Gamma-Ray Spectrum Analysis and MCA Emulator*, Software User's Manual (V. 5.1) Part No. 774780 (1999).

8. Beck, H. L., Planque, G., 1968. *The radiation field in air due to distributed gamma ray sources in ground*. HASL-195, Environmental Measurements Laboratory, New York: U.S. DOE, (1968).

9. Beck, H. L., DeCampo, J., Gogolak, C., 1972. *In situ Ge(Li) and NaI(Tl) gamma-ray spectrometry*. HASL-258, Environmental Measurements Laboratory, New York: U.S. DOE (1972).

10. Clouvas, A., Xanthos, S. and Antonopoulos-Domis, M. *Monte Carlo calculation of dose rate conversion factors for external exposure to photon emitters in soil*. Health Phys. **78** (3), 295-302 (2000).

11. Chen, S. Y. *Calculation of effective dose-equivalent responses for external exposure from residual photon emitters in soil*. Health Phys. **60**, 411-426 (1991).

12. Saito, K. and Jacob, P. *Gamma ray fields in the air due to sources in the ground*. Radiat. Prot.. Dosim. **58**, 29-45 (1995).

13. UNSCEAR. *Ionizing Radiation sources and biological effects*. A/37/45, United Nations, New York (1982).

14. Svoukis, E. and Tsertos, H. *Indoor and outdoor in situ high-resolution gamma radiation measurements in urban areas of Cyprus*. Radiat. Protect. Dosim. **123**(3), 384-390 (2007).

15. Anastasiou, T., Tsertos, H., Christofides, S. and Christodoulides, G., *Indoor radon concentration measu-rements in Cyprus using high sensitivity portable detectors*. J. Environ. Radioact. **68**, 159-169 (2003).

16. Nuclear Energy Agency. *Exposure to radiation from natural radioactivity in building materials*. Report by NEA Group of Experts, OECD: Paris, France (1979).

17. Beretka, J. and Mathew, P. J. *Natural radioactivity in Australian building materials, industrial waste and by-product*. Health Phys. **48**, 87-95 (1985).

18. Slunga, E. *Radon Classification of building ground*. Radiat. Prot. Dosim. **24**(114), 39-42 (1988).

19. Hayumbu, P., Zaman, M. B., Lubaba, N.C. H., Munsanje, S. S. and Nuleya, D. 1995. *Natural radioactivity in Zambian building materials collected from Lusaka*. J. Radioanal. Nucl. Chem. **199**, 229-238 (1995).

**Table 1. The activity concentrations of [226]Ra, [238]U, [232]Th, and [40]K in materials used for building purposes in Cyprus.**

| Sample Number | Raw and Rock Building Materials (origin) | Activity Concentrations | | | |
|---|---|---|---|---|---|
| | | [226]Ra | [238]U | [232]Th | [40]K |
| Quarry materials from Cyprus | | | | | |
| 1 | Six Diabase rocks (Stavrovouni)[2] | 0.75 ± 0.2 | 0.8 ± 0.1 | 0.3 ± 0.2 | 61.2 ± 2.9 |
| 2 | Three Diabase rocks (Pelendri)[2] | 0.1[1] | 1.1 ± 0.1 | 0.6 ± 0.3 | 138.3 ± 5.1 |
| 3 | Three Calcareous sandstones (Kellia)[2] | 40.2 ± 2.1 | 19.4 ± 0.5 | 3.8 ± 0.3 | 72.5 ± 2.9 |
| 4 | Three Diabase rocks (Sia)[2] | 1.2 ± 0.3 | 1.3 ± 0.2 | 0.7 ± 0.3 | 109.9 ± 4.0 |
| 5 | Calcareouslimestone (Sozomenos) | 30.9 ± 3.0 | 17.0 ± 0.6 | 6.0 ± 0.5 | 147.3 ± 5.4 |
| 6 | Four Yfalogenous limestones (Xylofagou)[2] | 7.0 ± 1.5 | 7.2 ± 0.3 | 0.6 ± 0.2 | 4.6 ± 1.5 |
| 7 | Two Diabase rocks (Vasa)[2] | 3.0 ± 0.7 | 1.3 ± 0.1 | 0.5 ± 0.2 | 113.3 ± 3.9 |
| 8 | Two Diabase rocks (Pareklissia)[2] | 3.3 ± 0.6 | 1.4 ± 0.2 | 1.0 ± 0.3 | 41.9 ± 2.5 |
| 9 | Sand (Mitsero) | 6.4 ± 2.0 | 5.7 ± 0.2 | 0.2 ± 0.3 | 40.5 ± 2.2 |
| 10 | Chalk (Kalo Xorio) | 33.8 ± 2.2 | 21.2 ± 0.7 | 6.4 ± 0.4 | 118.6 ± 4.8 |
| 11 | Diabase rock (Mosfiloti) | 4.4 ± 1.0 | 1.7 ± 0.2 | 1.1 ± 0.2 | 69.1 ± 2.5 |
| 12 | Limestone (Androlykou) | 20.9 ± 1.5 | 8.3 ± 0.3 | 0.3 ± 0.2 | 6.6 ± 1.0 |
| 13 | Diadase rock (Pyrga) | 5.3 ± 1.0 | 0.9 ± 0.1 | 0.5 ± 0.2 | 55.8 ± 2.2 |
| A.M. ± S.D.[3] | | 12.1 ± 14.1 | 6.7 ± 7.6 | 1.7 ± 2.2 | 75.4 ± 46.9 |
| Cements from Cyprus | | | | | |
| 14 | EN 197-1 Cem I 42.5 R | 22.9 ± 3.2 | 13.3 ± 0.6 | 9.2 ± 0.5 | 160.9 ± 5.9 |
| 15 | EN 197-I Cem I 52.5 N | 4.4 ± 1.5 | 9.3 ± 0.4 | 7.7 ± 0.4 | 91.8 ± 3.8 |
| 16 | EN 197-I Cem II A/L 42.5 N | 24.1 ± 2.7 | 15.2 ± 0.5 | 4.9 ± 0.4 | 4.1 ± 1.6 |
| 17 | EN197-I Cem II/A-M(L-S)42.5r | 36.6 ± 3.4 | 20.7 ± 0.7 | 12.3 ± 0.5 | 208.9 ± 7.3 |
| 18 | EN 197-I Cem II/A-M(L-S)52.5 | 33.1 ± 2.3 | 18.4 ± 0.6 | 10.4 ± 0.5 | 150.8 ± 5.8 |
| 19 | EN 197-I CemII/A-P42.5N | 25.4 ± 3.0 | 16.4 ± 0.5 | 11.4 ± 0.6 | 194.4 ± 6.7 |
| 20 | EN 197-I CemII/B-M(L-S)32.5r | 20.4 ± 3.3 | 13.1 ± 0.5 | 9.6 ± 0.5 | 207.1 ± 7.2 |
| 21 | EN 197-I Cem II /A-P 42.5N | 31.6 ± 3.6 | 19.0 ± 0.6 | 12.1 ± 0.5 | 199.2 ± 7.0 |
| 22 | Gypsos (Toxni) | 60.3 ± 2.8 | 24.5 ± 0.7 | 2.9 ± 0.3 | 40.1 ± 2.1 |
| 23 | Chalk (Kalavasos) | 12.0 ± 1.3 | 3.8 ± 0.2 | 6.8 ± 0.4 | 92.2 ± 3.2 |
| 24 | Limestone (Armenoxori) | 37.0 ± 1.7 | 16.8 ± 0.5 | 0.6 ± 0.2 | 8.1 ± 1.2 |
| 25 | Marl (Vasiliko) | 47.6 ± 2.7 | 17.5 ± 0.6 | 6.5 ± 0.4 | 289.2 ± 8.1 |
| 26 | Cement Tile Mantonella-Lydia | 12.0 ± 2.1 | 8.6 ± 0.3 | 1.7 ± 0.3 | 4.1 ± 1.5 |
| A.M. ± S.D. | | 28.3 ± 15.3 | 15.1 ± 5.5 | 7.4 ± 3.9 | 127.0 ± 93.7 |
| Clays from Cyprus | | | | | |
| 27 | Two Red Clay (Kornos)[2] | 0.1[1] | 2.7 ± 0.3 | 2.5 ± 0.3 | 65.6 ± 3.2 |
| 28 | Grey Clay (Tseri/Ayia Varnara) | 0.2[1] | 13.1 ± 0.5 | 5.7 ± 0.4 | 292.5 ± 10.5 |
| 29 | Mixture of Red (Lythrodontas) and Grey Clay (Tseri/Levkara) | 0.2[1] | 6.7 ± 0.3 | 4.5 ± 0.4 | 192.3 ± 7.2 |
| 30 | Two Grey Clay (Tseri)[2] | 12.6 ± 2.6 | 10.1 ± 0.4 | 3.7 ± 0.5 | 229.9 ± 8.0 |
| 31 | Red Clay (Levkara) | 0.1[1] | 2.4 ± 0.2 | 1.8 ± 0.3 | 59.3 ± 3.0 |
| 32 | Grey Clay (Pera Xorio/Nisou) | 27.8 ± 2.6 | 15.0 ± 0.5 | 8.0 ± 0.6 | 320.6 ± 10.9 |
| 33 | Red Clay (Xylofagou) for tiles | 33.8 ± 3.8 | 20.6 ± 0.8 | 16.0 ± 0.7 | 376.9 ± 12.2 |
| 34 | Mixture of Red (Kornos) and Grey (Tseri) Clay for bricks | 3.5 ± 1.6 | 8.0 ± 0.3 | 4.0 ± 0.3 | 165.1 ± 6.1 |
| 35 | Red Clay(Levkara/Lythrodontas) | 14.0 ± 1.3 | 3.9 ± 0.2 | 4.4 ± 0.3 | 102.2 ± 3.0 |
| A.M. ± S.D. | | 10.3 ± 12.9 | 9.2 ± 6.2 | 5.6 ± 4.3 | 200.5 ± 114. |
| Imported Granites | | | | | |
| 36 | Chiva Cashi (Italy) | 0.3[1] | 106. ± 3.0 | 5.7 ± 0.7 | 1143.5 ± 33.0 |
| 37 | Lambrileta (Italy) | 2.2[1] | 22.8 ± 0.7 | 209. ± 6.0 | 1576.0 ± 45.2 |
| 38 | Ligth Pink (China) | 0.2[1] | 25.1 ± 0.8 | 43.1 ± 1.3 | 920.7 ± 26.8 |
| 39 | All white fine (Italy) | 4.6 ± 1.2 | 3.2 ± 0.2 | 2.4 ± 0.3 | 349.9 ± 9.1 |
| 40 | Bianco Perla (Italy | 215. ± 6.3 | 98.9 ± 2.1 | 127. ± 2.8 | 1360.6 ± 29.1 |
| 41 | Santa Cechilia (Italy) | 36.0 ± 2.2 | 18.0 ± 0.5 | 45.9 ± 1.1 | 1353.1 ± 28.9 |
| 42 | Yellow Light (Italy) | 139. ± 4.7 | 52.9 ± 1.4 | 45.4 ± 1.2 | 1213.0 ± 30.0 |
| 43 | Giallo California (Italy) | 81.0 ± 3.8 | 107. ± 2.6 | 260. ± 6.3 | 1212.9 ± 29.6 |
| A.M. ± S.D. | | 59.8 ± 80.0 | 54.2 ± 43.4 | 92.3 ± 96.6 | 1141.2 ± 372. |
| Imported Marbles | | | | | |
| 44 | Emarator Light (Italy) | 41.6 ± 2.8 | 24.0 ± 0.7 | 0.5 ± 0.3 | 65.6 ± 2.8 |
| 45 | Karnazeiko (Greece) | 0.1[1] | 2.8 ± 0.1 | 0.2 ± 0.1 | 3.9 ± 1.2 |
| 46 | Verde Venado (Italy) | 2.1 ± 1.3 | 3.4 ± 0.2 | 0.8 ± 0.1 | 23.3 ± 1.8 |
| 47 | Lambrileta (Italy) | 7.9 ± 1.0 | 2.1 ± 0.1 | 0.7 ± 0.2 | 4.6 ± 0.9 |
| 48 | Perlato Royal (Italy) | 5.8 ± 1.0 | 1.3 ± 0.1 | 0.1[1] | 0.4[1] |
| 49 | Pirgon (Italy) | 63.7 ± 2.2 | 33.6 ± 0.8 | 0.1[1] | 0.2[1] |

| | | | | | | | | | |
|---|---|---|---|---|---|---|---|---|---|
| 50 | Saint Lorent (Italy) | 44.0 | ± 1.7 | 27.0 | ± 0.7 | 0.1[1] | | 0.2[1] | |
| 51 | Travertino (Italy) | 79.0 | ± 2.6 | 35.3 | ± 0.8 | 0.9 | ± 0.2 | 15.1 | ± 1.1 |
| 52 | Travertino Giallo (Italy) | 1.8 | ± 1.2 | 7.2 | ± 0.3 | 2.1 | ± 0.2 | 26.8 | ± 1.4 |
| 53 | Civec (Greece) | 9.8 | ± 1.0 | 5.5 | ± 0.2 | 0.9 | ± 0.2 | 16.3 | ± 1.3 |
| A.M. ± S.D. | | 25.6 | ± 29.1 | 14.2 | ± 14.0 | 0.64 | ± 0.61 | 15.6 | ± 20.1 |
| Imported Ceramics | | | | | | | | | |
| 54 | Safari (Italy) | 83.2 | ± 3.3 | 8.6 | ± 0.3 | 1.7 | ± 0.3 | 4.1 | ± 1.5 |
| 55 | Marble Slip Rojo (Spain) | 137. | ± 5.0 | 50.7 | ± 1.3 | 59.7 | ± 1.5 | 573.8 | ± 14.5 |
| 56 | Gomez (Spain) | 141 | ± 5.5 | 67.1 | ± 1.7 | 67.6 | ± 1.7 | 935.7 | ± 23.1 |
| 57 | Purhase Beige (Spain) | 30.1 | ± 2.3 | 68.1 | ± 1.7 | 74.6 | ± 1.9 | 1051.8 | ± 26.0 |
| 58 | Colorado (Spain) | 117. | ± 3.7 | 66.1 | ± 1.7 | 63.2 | ± 1.7 | 783.0 | ± 19.6 |
| 59 | Castelveto T1 (Italy) | 28.9 | ± 2.8 | 64.9 | ± 1.6 | 69.5 | ± 1.8 | 389.2 | ± 10.0 |
| 60 | 3965 (Spain) | 167. | ± 5.6 | 64.9 | ± 1.6 | 54.8 | ± 1.5 | 771.9 | ± 19.5 |
| 61 | 6LSP004 (China) | 281. | ± 8.2 | 154. | ± 3.8 | 141. | ± 3.5 | 623.0 | ± 15.8 |
| 62 | Castelvetro T2 (Italy) | 130. | ± 4.8 | 136. | ± 3.3 | 184. | ± 4.5 | 485.7 | ± 12.3 |
| 63 | Project Grey (Italy) | 299. | ± 8.8 | 66. | ± 1.7 | 58.0 | ± 1.5 | 877.3 | ± 22.0 |
| 64 | Pergamon (Italy) | 97.6 | ± 3.6 | 140. | ± 3.4 | 208. | ± 5.1 | 504.9 | ± 13.0 |
| 65 | Oscar Oro (China) | 105. | ± 3.8 | 45.4 | ± 1.2 | 51.2 | ± 1.4 | 495.5 | ± 12.5 |
| 66 | Mosca (Spain) | 127. | ± 4.1 | 51.4 | ± 1.3 | 59.0 | ± 1.6 | 564.8 | ± 14.3 |
| A.M. ± S.D. | | 134.1 | ± 80.0 | 75.6 | ± 41.9 | 84.0 | ± 57.9 | 620.0 | ± 272. |
| Decorative Stones | | | | | | | | | |
| 67 | Slide (Italy) | 53.3 | ± 2.7 | 20.1 | ± 0.5 | 50.8 | ± 1.2 | 1010.4 | ± 21.7 |
| 68 | Beige (Cyprus) | 23.4 | ± 2.3 | 14.3 | ± 0.5 | 1.1 | ± 0.3 | 5.8 | ± 1.5 |
| 69 | Toxnis (Cyprus) | 42.3 | ± 2.3 | 19.6 | ± 0.5 | 2.4 | ± 0.2 | 21.4 | ± 1.3 |
| A.M. ± S.D. | | 39.7 | ± 15.1 | 18.0 | ± 3.2 | 18.1 | ± 28.3 | 345.8 | ± 576. |

**Table 2. The calculated total absorbed dose rates (D), the annual outdoor and indoor effective dose rates ($D_E$), the radium equivalent activity ($Ra_{eq}$), the external hazard index ($H_{ex}$), and the activity concentration index (I), for the corresponding measured samples of Table 1.**

| | Sample | D | | $D_{E,\ out}$ | | $D_{E,\ in}$ | | $Ra_{eq}$ | | $H_{ex}$ | | I | |
|---|---|---|---|---|---|---|---|---|---|---|---|---|---|
| Quarry Materials | 1 | 2.8 | ±0.2 | 3.5 | ±0.2 | 19.5 | ±0.4 | 5.9 | ±0.6 | 0.016 | ±0.001 | 0.024 | ±0.002 |
| | 2 | 6.1 | ±0.3 | 7.5 | ±0.3 | 41.8 | ±0.6 | 11.6 | ±0.8 | 0.031 | ±0.002 | 0.049 | ±0.002 |
| | 3 | 12.4 | ±0.3 | 15.2 | ±0.3 | 84.9 | ±0.7 | 51.2 | ±2.2 | 0.138 | ±0.006 | 0.177 | ±0.007 |
| | 4 | 5.1 | ±0.2 | 6.3 | ±0.3 | 35.2 | ±0.6 | 10.7 | ±0.8 | 0.029 | ±0.002 | 0.044 | ±0.002 |
| | 5 | 15.5 | ±0.4 | 19.0 | ±0.5 | 106.3 | ±1.0 | 50.8 | ±3.2 | 0.137 | ±0.008 | 0.182 | ±0.010 |
| | 6 | 3.3 | ±0.2 | 4.0 | ±0.2 | 22.6 | ±0.4 | 8.2 | ±1.6 | 0.022 | ±0.004 | 0.028 | ±0.005 |
| | 7 | 5.1 | ±0.2 | 6.3 | ±0.2 | 35.3 | ±0.5 | 12.4 | ±0.9 | 0.034 | ±0.002 | 0.050 | ±0.003 |
| | 8 | 2.7 | ±0.2 | 3.3 | ±0.2 | 18.5 | ±0.5 | 8.0 | ±0.9 | 0.021 | ±0.002 | 0.030 | ±0.003 |
| | 9 | 3.9 | ±0.2 | 4.8 | ±0.2 | 26.7 | ±0.5 | 9.8 | ±2.1 | 0.026 | ±0.006 | 0.036 | ±0.007 |
| | 10 | 16.2 | ±0.4 | 19.9 | ±0.5 | 111.3 | ±1.0 | 52.1 | ±2.4 | 0.141 | ±0.006 | 0.184 | ±0.008 |
| | 11 | 3.9 | ±0.2 | 4.8 | ±0.2 | 26.9 | ±0.4 | 11.3 | ±1.1 | 0.031 | ±0.003 | 0.043 | ±0.004 |
| | 12 | 3.6 | ±0.2 | 4.5 | ±0.2 | 25.0 | ±0.4 | 21.8 | ±1.6 | 0.059 | ±0.004 | 0.073 | ±0.005 |
| | 13 | 2.8 | ±0.1 | 3.4 | ±0.2 | 19.0 | ±0.3 | 10.3 | ±1.1 | 0.028 | ±0.003 | 0.039 | ±0.004 |
| Cements | 14 | 16.2 | ±0.4 | 19.9 | ±0.5 | 111.6 | ±1.0 | 48.4 | ±3.3 | 0.131 | ±0.009 | 0.176 | ±0.011 |
| | 15 | 11.2 | ±0.3 | 13.8 | ±0.4 | 77.1 | ±0.7 | 22.5 | ±1.7 | 0.061 | ±0.004 | 0.084 | ±0.006 |
| | 16 | 8.7 | ±0.3 | 10.6 | ±0.3 | 59.5 | ±0.7 | 31.4 | ±2.8 | 0.085 | ±0.007 | 0.106 | ±0.009 |
| | 17 | 22.6 | ±0.5 | 27.7 | ±0.6 | 155.3 | ±1.2 | 70.3 | ±3.6 | 0.190 | ±0.010 | 0.253 | ±0.012 |
| | 18 | 18.5 | ±0.4 | 22.7 | ±0.5 | 126.9 | ±1.0 | 59.6 | ±2.5 | 0.161 | ±0.007 | 0.213 | ±0.008 |
| | 19 | 19.9 | ±0.5 | 24.4 | ±0.6 | 136.7 | ±1.1 | 56.7 | ±3.2 | 0.153 | ±0.009 | 0.206 | ±0.011 |
| | 20 | 18.2 | ±0.4 | 22.3 | ±0.5 | 124.8 | ±1.1 | 50.1 | ±3.5 | 0.135 | ±0.009 | 0.185 | ±0.012 |
| | 21 | 21.5 | ±0.4 | 26.3 | ±0.5 | 147.5 | ±1.1 | 64.2 | ±3.7 | 0.174 | ±0.010 | 0.232 | ±0.012 |
| | 22 | 12.6 | ±0.3 | 15.5 | ±0.4 | 86.6 | ±0.8 | 67.5 | ±2.9 | 0.183 | ±0.008 | 0.229 | ±0.009 |
| | 23 | 8.6 | ±0.3 | 10.6 | ±0.3 | 59.3 | ±0.6 | 28.8 | ±1.5 | 0.078 | ±0.004 | 0.105 | ±0.005 |
| | 24 | 7.2 | ±0.2 | 8.8 | ±0.3 | 49.2 | ±0.6 | 38.5 | ±1.8 | 0.104 | ±0.005 | 0.129 | ±0.006 |
| | 25 | 21.4 | ±0.4 | 26.3 | ±0.5 | 147.0 | ±1.1 | 79.2 | ±2.9 | 0.214 | ±0.008 | 0.288 | ±0.010 |
| | 26 | 4.4 | ±0.2 | 5.4 | ±0.3 | 30.2 | ±0.5 | 14.7 | ±2.2 | 0.040 | ±0.006 | 0.050 | ±0.007 |
| Clays | 27 | 4.9 | ±0.2 | 6.0 | ±0.3 | 33.7 | ±0.6 | 8.7 | ±0.7 | 0.024 | ±0.001 | 0.035 | ±0.002 |
| | 28 | 19.4 | ±0.6 | 23.8 | ±0.7 | 133.3 | ±1.4 | 30.9 | ±1.3 | 0.083 | ±0.003 | 0.127 | ±0.005 |
| | 29 | 12.4 | ±0.4 | 15.2 | ±0.5 | 85.2 | ±0.9 | 21.4 | ±0.9 | 0.058 | ±0.002 | 0.087 | ±0.003 |
| | 30 | 14.8 | ±0.4 | 18.1 | ±0.5 | 101.4 | ±1.1 | 35.6 | ±2.8 | 0.096 | ±0.007 | 0.137 | ±0.009 |
| | 31 | 4.2 | ±0.2 | 5.1 | ±0.3 | 28.7 | ±0.5 | 7.2 | ±0.6 | 0.020 | ±0.001 | 0.029 | ±0.002 |
| | 32 | 22.4 | ±0.6 | 27.5 | ±0.7 | 154.1 | ±1.4 | 63.9 | ±2.9 | 0.173 | ±0.008 | 0.240 | ±0.010 |
| | 33 | 31.0 | ±0.7 | 38.0 | ±0.8 | 213.0 | ±1.7 | 85.7 | ±4.0 | 0.231 | ±0.011 | 0.318 | ±0.014 |
| | 34 | 11.6 | ±0.3 | 14.2 | ±0.4 | 79.7 | ±0.8 | 21.9 | ±1.8 | 0.059 | ±0.005 | 0.087 | ±0.006 |
| | 35 | 7.8 | ±0.2 | 9.6 | ±0.3 | 53.5 | ±0.5 | 28.2 | ±1.5 | 0.076 | ±0.004 | 0.103 | ±0.005 |
| Granites | 36 | 88.4 | ±1.8 | 108.4 | ±2.2 | 607.2 | ±4.3 | 96.5 | ±1.9 | 0.261 | ±0.007 | 0.411 | ±0.012 |
| | 37 | 180.1 | ±3.6 | 220.9 | ±4.4 | 1237. | ±8.9 | 422.4 | ±9.5 | 1.141 | ±0.027 | 1.578 | ±0.036 |
| | 38 | 68.1 | ±1.3 | 83.5 | ±1.6 | 467.6 | ±3.1 | 132.7 | ±2.3 | 0.358 | ±0.007 | 0.523 | ±0.011 |
| | 39 | 16.0 | ±0.4 | 19.7 | ±0.5 | 110.0 | ±1.0 | 35.0 | ±1.5 | 0.094 | ±0.004 | 0.144 | ±0.005 |
| | 40 | 158.1 | ±2.0 | 193.9 | ±2.5 | 1086. | ±5.0 | 501.4 | ±7.6 | 1.354 | ±0.021 | 1.805 | ±0.027 |
| | 41 | 83.5 | ±1.3 | 102.4 | ±1.6 | 573.4 | ±3.1 | 205.8 | ±3.1 | 0.556 | ±0.009 | 0.801 | ±0.013 |
| | 42 | 91.4 | ±1.4 | 112.1 | ±1.8 | 627.7 | ±3.5 | 297.3 | ±5.2 | 0.803 | ±0.015 | 1.095 | ±0.020 |
| | 43 | 225.8 | ±3.7 | 276.9 | ±4.5 | 1551. | ±9.0 | 546.2 | ±9.9 | 1.475 | ±0.027 | 1.974 | ±0.035 |
| Marbles | 44 | 12.1 | ±0.3 | 14.9 | ±0.4 | 83.4 | ±0.8 | 47.4 | ±2.9 | 0.128 | ±0.008 | 0.163 | ±0.009 |
| | 45 | 1.3 | ±0.1 | 1.7 | ±0.1 | 9.2 | ±0.3 | 0.7 | ±0.4 | 0.002 | ±0.001 | 0.003 | ±0.001 |
| | 46 | 2.6 | ±0.1 | 3.2 | ±0.1 | 18.2 | ±0.3 | 5.0 | ±1.4 | 0.014 | ±0.004 | 0.019 | ±0.004 |
| | 47 | 1.4 | ±0.1 | 1.7 | ±0.1 | 9.4 | ±0.3 | 9.3 | ±1.1 | 0.025 | ±0.003 | 0.031 | ±0.003 |
| | 48 | 0.6 | ±0.1 | 0.7 | ±0.1 | 3.9 | ±0.3 | 6.0 | ±1.0 | 0.016 | ±0.003 | 0.020 | ±0.003 |
| | 49 | 13.1 | ±0.3 | 16.0 | ±0.4 | 89.9 | ±0.8 | 63.7 | ±2.2 | 0.172 | ±0.006 | 0.212 | ±0.007 |
| | 50 | 10.6 | ±0.3 | 13.0 | ±0.3 | 72.6 | ±0.7 | 44.2 | ±1.7 | 0.119 | ±0.005 | 0.147 | ±0.006 |
| | 51 | 14.8 | ±0.3 | 18.1 | ±0.4 | 101.6 | ±0.8 | 81.5 | ±2.6 | 0.220 | ±0.007 | 0.273 | ±0.009 |
| | 52 | 4.9 | ±0.2 | 6.1 | ±0.2 | 34.0 | ±0.4 | 6.9 | ±1.3 | 0.019 | ±0.003 | 0.025 | ±0.004 |
| | 53 | 3.2 | ±0.1 | 4.0 | ±0.2 | 22.3 | ±0.3 | 12.3 | ±1.1 | 0.033 | ±0.003 | 0.043 | ±0.004 |
| Ceramics | 54 | 4.4 | ±0.2 | 5.4 | ±0.3 | 30.2 | ±0.5 | 85.9 | ±3.3 | 0.232 | ±0.009 | 0.287 | ±0.011 |
| | 55 | 73.4 | ±1.1 | 90.0 | ±1.3 | 504.2 | ±2.7 | 266.6 | ±5.5 | 0.720 | ±0.015 | 0.946 | ±0.019 |
| | 56 | 97.9 | ±1.4 | 120.1 | ±1.8 | 672.7 | ±3.5 | 309.7 | ±6.2 | 0.837 | ±0.017 | 1.120 | ±0.022 |
| | 57 | 106.5 | ±1.6 | 130.6 | ±1.9 | 731.5 | ±3.8 | 217.8 | ±3.8 | 0.588 | ±0.011 | 0.824 | ±0.015 |
| | 58 | 89.3 | ±1.3 | 109.6 | ±1.7 | 613.5 | ±3.3 | 267.7 | ±4.6 | 0.723 | ±0.013 | 0.967 | ±0.016 |
| | 59 | 77.0 | ±1.2 | 94.4 | ±1.5 | 528.8 | ±2.9 | 158.3 | ±3.9 | 0.427 | ±0.010 | 0.574 | ±0.013 |
| | 60 | 84.0 | ±1.3 | 103.0 | ±1.5 | 576.9 | ±3.1 | 304.8 | ±6.1 | 0.823 | ±0.017 | 1.088 | ±0.021 |
| | 61 | 158.5 | ±2.4 | 194.3 | ±2.9 | 1088. | ±5.8 | 530.2 | ±9.6 | 1.432 | ±0.026 | 1.848 | ±0.032 |
| | 62 | 168.9 | ±2.7 | 207.1 | ±3.4 | 1160. | ±6.7 | 430.2 | ±8.1 | 1.162 | ±0.022 | 1.514 | ±0.028 |
| | 63 | 90.2 | ±1.3 | 110.6 | ±1.6 | 619.4 | ±3.3 | 450.0 | ±9.2 | 1.216 | ±0.025 | 1.581 | ±0.031 |
| | 64 | 183.8 | ±3.0 | 225.5 | ±3.7 | 1262. | ±7.5 | 433.9 | ±8.2 | 1.172 | ±0.022 | 1.534 | ±0.029 |
| | 65 | 63.8 | ± 1.0 | 78.3 | ± 1.2 | 438.4 | ± 2.5 | 216.9 | ± 4.4 | 0.586 | ± 0.012 | 0.773 | ± 0.015 |
| | 66 | 73.0 | ± 1.1 | 89.5 | ± 1.4 | 501.2 | ± 2.8 | 255.3 | ± 4.8 | 0.690 | ± 0.013 | 0.908 | ± 0.017 |
| Stones | 67 | 73.7 | ± 1.1 | 90.3 | ± 1.3 | 505.9 | ± 2.6 | 203.7 | ± 3.5 | 0.550 | ± 0.010 | 0.768 | ± 0.013 |
| | 68 | 6.4 | ± 0.3 | 7.8 | ± 0.3 | 43.8 | ± 0.6 | 25.4 | ± 2.4 | 0.069 | ± 0.006 | 0.085 | ± 0.008 |
| | 69 | 9.7 | ± 0.2 | 11.9 | ± 0.3 | 66.8 | ± 0.6 | 47.4 | ± 2.3 | 0.128 | ± 0.006 | 0.160 | ± 0.008 |

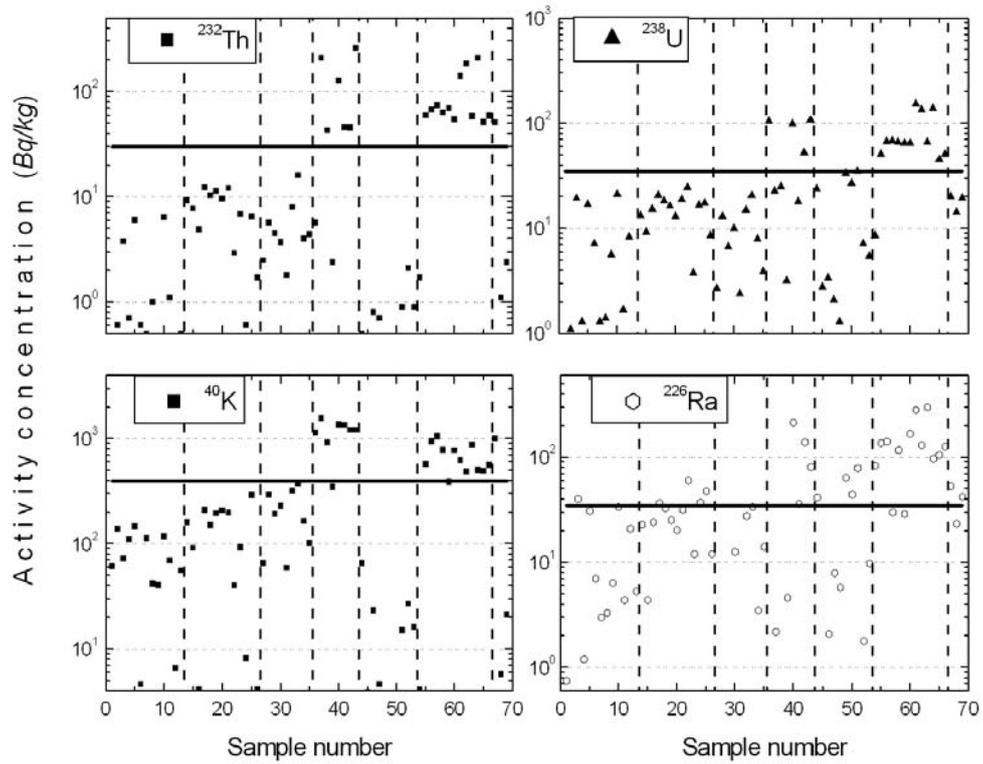

*Figure 1. The $A_{Ra}$, $A_U$, $A_{Th}$, $A_K$ of the measured samples. The horizontal solid line in each graph corresponds to the world [1] median radioactivity levels ($A_{Ra} = 35$ Bq kg⁻¹, $A_U = 35$ Bq kg⁻¹, $A_{Th} = 30$ Bq kg⁻¹, $A_K = 400$ Bq kg⁻¹). The vertical dashed lines in each graph correspond to the seven sample categories in Table 1.*

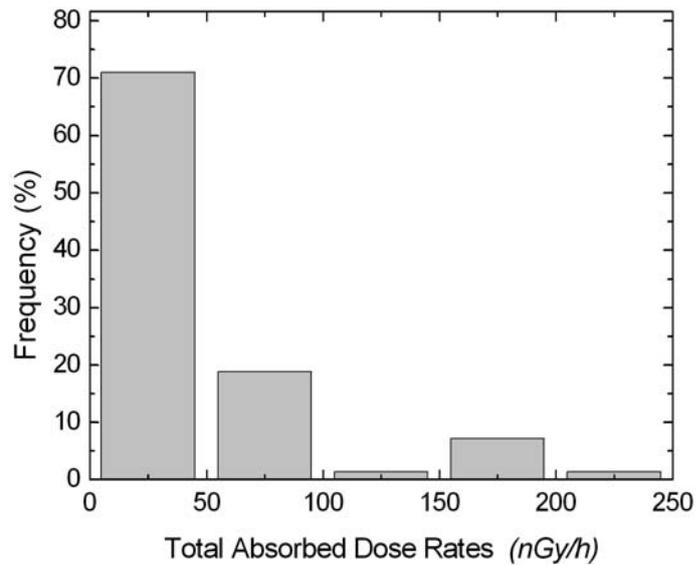

*Figure 2. The percentage frequency distribution of the total absorbed rates from the measured samples in Table 1.*

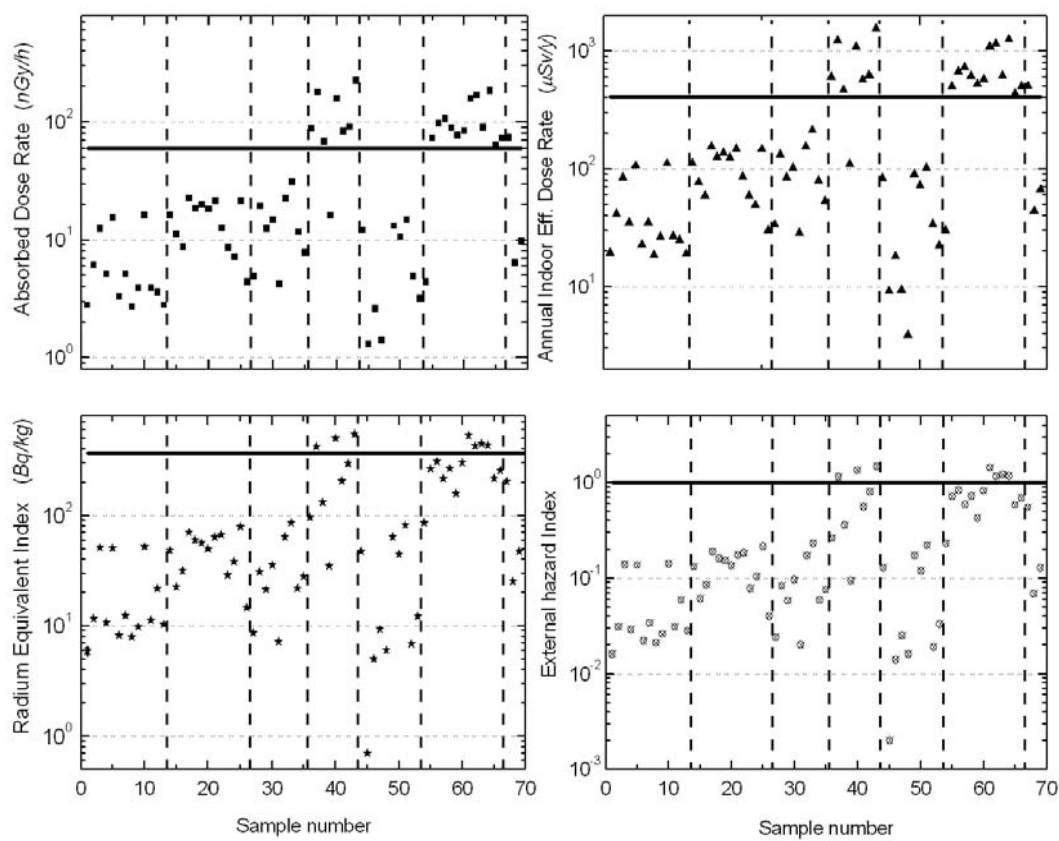

*Figure 3. The calculated absorbed dose rates (D), the annual indoor effective dose rates ($D_{E, in}$), the radium equivalent activity ($Ra_{eq}$), and the external hazard index ($H_{ex}$) for the corresponding measured samples of Table 1. The horizontal solid lines correspond to D = 60 nGy h[-1] and $D_{E, in}$ = 410 µSv y[-1] world median values [1], respectively, and to the $Ra_{eq}$ = 370 Bq kg[-1] and $H_{ex}$ = 1 world upper limit values of radiation dose from building materials to the inhabitants[13,18], respectively. The vertical dashed lines in each graph correspond to the seven sample categories in Table 1.*